\documentclass[onecolumn, showpacs, aps, superscriptaddress, amsmath, amssymb, nofootinbib]{revtex4-2}
\usepackage[T1]{fontenc}
\usepackage{notes2bib}
\usepackage{lmodern}
\usepackage{graphicx}
\usepackage{upgreek}
\usepackage[mathscr]{eucal}
\usepackage{amsmath}
\usepackage{amssymb}

\usepackage{bm}
\usepackage{color}
\usepackage[dvipsnames]{xcolor}
\usepackage{hyperref}
\hypersetup{colorlinks=true,citecolor=blue}
\usepackage{comment}
\usepackage{cleveref}
\usepackage{multirow}
\usepackage[percent]{overpic}

\crefname{equation}{Eq.}{Eqs.}
\allowdisplaybreaks

\usepackage{tikz}

\newcommand{\customref}[2]{\hyperref[#1]{#2}}

\begin{document}

\title{Rheotaxis of microswimmers in colloid-laden channel flow}

\author{Margam Ramprasad}
\affiliation{Department of Mechanical Engineering, Indian Institute of Technology, Madras, Chennai 600036, India}
\author{Shubhadeep Mandal}
\email{smandal@iisc.ac.in}
\affiliation{Department of Mechanical Engineering, Indian Institute of Science, Bengaluru 560012, India}
\author{Pallab Sinha Mahapatra}
\email{pallab@iitm.ac.in}
\affiliation{Department of Mechanical Engineering, Indian Institute of Technology, Madras, Chennai 600036, India}

\date{\today}

\begin{abstract}
Microswimmers are often found in heterogeneous and crowded environments within narrow conduits under external flow conditions, enabling them to perform interesting translational and rotational maneuvers, such as swimming in the upstream direction, following walls, and oscillatory motion. Studying such systems helps us understand the motility behaviors of microswimmers (pushers, pullers, or neutrals) and develop applications such as targeted drug delivery. To study the motion of microswimmers in a channel flow with the presence of hard, monodisperse spherical colloids, we adopted the spherical squirmer model to represent the microswimmers, along with a mesoscale simulation framework, multi-particle collision dynamics (MPCD), to represent the background fluid. In the absence of colloids, a squirmer in a microchannel flow develops an increased probability of moving away from the walls and oscillates between the walls as the flow speed increases compared to the squirmer speed, with a dominant upstream orientation near the walls. However, the presence of the colloids makes the pusher swim towards the center of the channel and upstream direction, and the puller swim away from the center of the channel at low flow speeds. At high flow speeds, the flow carries all the squirmers, resulting in a dominant upstream direction in the channel center. We observe that this leads to a decrease in the local velocity of the squirmer in the flow direction for pusher, neutral, and puller-type squirmers. We also observe that, for a constant colloidal packing fraction, the local velocity magnitude of the puller along the flow direction is less than that of the pusher.
\end{abstract}


\maketitle

\section{\label{sec:Intro}Introduction}

 Natural microswimmers like motile bacteria, algae, and sperm cells are often found in heterogeneous and crowded environments \cite{Lauga2016bacteria, Aragones2018, Conrad2018, Spagnolie2023}, following which artificial microswimmers like active Janus particles, active droplets, and microrobots \cite{Bechinger2016, Simmchen2016janus, Bunea2020}, are developed. In such systems, the microswimmers move their flagella or cilia in such a way that they break the time-reversal symmetry to aid in the self-propulsion \cite{Purcell1977lowRe}. Based on the velocity profile generated by the microswimmers, they are segregated into three types: pusher, neutral, and puller. The pushers generate the thrust force from the back, whereas the pullers generate the thrust force in the front to move forward \cite{Gotze2010}. Similarly, the neutral type microswimmers generate net zero thrust on the fluid to move forward. It is known that the motion of the microswimmers is affected by the background fluid \cite{Schneider1973enhancement, Zottl2019}, the surrounding medium, like colloids \cite{Kamdar2022} and walls \cite{Schaar2015wall, Theers2016wall}, and external stimuli \cite{Pedley1992, Ishikawa2025TransportRheology} like chemical gradient for chemotaxis \cite{Berg1977, Bhattacharjee2021chemotaxis}, magnetic field for magnetotaxis \cite{Tierno2008MagneticallyMicroswimmers, Waisbord2021}, velocity gradient for rheotaxis \cite{Kaya2009flow, Zottl2012, Qi2020rheotaxis}, etc. Understanding these effects will provide insights, like targeted drug delivery, where the bioinspired microbots would swim through the bloodstream \cite{Nelson2010application, Wu2020review}, bacterial migration in wet soil and aquatic environments \cite{Hill2007experiment}, etc. \cite{DeSouza2015application, Thornlow2015application, Bunea2020, Ouyang2023}. To develop such applications, research is conducted on different microswimmer systems, like microswimmers in Newtonian fluids \cite{Lauga2009TheMicroorganisms, Ouyang2023}, viscoelastic fluids \cite{Kowalik2013, Espinosa-Garcia2013, Sahoo2019EnhancedFluid}, colloidal suspensions \cite{Leptos2009, Ji2011colloids, Ortlieb2019, Purushothaman2021}, polymer suspensions \cite{Winkler2004, Zottl2019, Zottl2023}, and structured obstacles \cite{Chopra2022, Ramprasad2025}, both experimentally and numerically.

Analysis of the motion of microswimmers in colloid-laden flow is one of such interesting topics, providing insights into the rheotaxis of microswimmers in heterogeneous environments. A simple example of rheotaxis is the controlled swimming of fish in the water stream \cite{Arnold1974}. In microscale environments, spermatozoa, bacteria \cite{Hill2007experiment, Kaya2009flow, Marcos2012BacterialRheotaxis, Rusconi2014}, and \textit{Chlamydomonas} \cite{Barry2015chlamydomonasrheo, Omori2022chlaydomonasrheo} exhibit rheotaxis. Initially, Bretherton and Rothschild \cite{Bretherton1961} reported the rheotaxis of spermatozoa, following which Kantsler et al. \cite{Kantsler2014} demonstrated the stable spiraling upstream motion of sperm cells. Hill et al. \cite{Hill2007experiment} observed that the \textit{E. coli} in a shear flow swim along the wall in the upstream direction over a long time range. Macros et al. \cite{Marcos2012BacterialRheotaxis} experimentally identified that bacteria also exhibit rheotaxis in the absence of a nearby surface. Rusconi et al. \cite{Rusconi2014} demonstrated the trapping of the bacteria (pusher) near the walls, whereas Barry et al. \cite{Barry2015chlamydomonasrheo} showed the \textit{Chlamydomonas} (puller) accumulation near the center of the channel. Omari et al. \cite{Omori2022chlaydomonasrheo} explained the upstream oriented swimming of \textit{Chlamydomonas} both experimentally and numerically. In analogy with microorganisms, synthetic microswimmers such as Janus particles, active droplets, and microrobots \cite{Ren2017microrobotrheo} also exhibit rheotaxis. Si et al. \cite{Si2020} observed the cross migration of Janus particles as the shear flow rate increases. Sharan et al. \cite{Sharan2022JPrheo} hypothesized that the Janus particles act like a puller and demonstrated the steady upstream motion. Dey et al. \cite{Dey2022OscillatoryMicrochannels} showed that the active droplets swim in the upstream direction in Poiseuille flow, and move in an oscillatory trajectory.

We find a wide range of numerical analyses that replicate the similar rheotaxis behaviors of the microswimmers \cite{Elgeti2015PhysicsReview, Ishimoto2023review}. In this context, Zöttl and Stark \cite{Zottl2012} investigated the nonlinear dynamics of a microswimmer swimming in the cylindrical Poiseuille flow, explaining the stable and unstable fixed points of the squirmer position and orientation with respect to the flow strength. Qi et al. \cite{Qi2020rheotaxis} studied the rheotaxis of the squirmer in a microchannel flow, explaining the effect of the flow strength and shape of the squirmer. Daddi-Moussa-Ider et al. \cite{Daddi2020} illustrated the controlled navigation of the microrobots with payloads using a three-sphere microswimmer developed by Najafi and Golestanian \cite{Najafi2004}. Supporting the current study, we find a wide range of research conducted in colloidal suspensions \cite{Winkler2005, Padding2006, Ji2011colloids}, the microswimmers in a quiescent heterogeneous medium \cite{Leptos2009, Molina2014, Chamolly2017, Aragones2018, Delmotte2018, Bhattacharjee2019}, and the motion of a microswimmer near a wall \cite{Theers2016wall, Kuron2019, Radhakrishnan2024}.

In the context of microswimmers swimming in quiescent heterogeneous media, we identify significant research both experimentally and numerically \cite{Patteson2016ActiveFluids, Shum2017, Wu2020review}. Molina et al. \cite{Molina2014} numerically studied the diffusion of the inert colloidal suspension in the presence of squirmers. Chamolly et al. \cite{Chamolly2017} performed numerical simulations to understand the translational behavior of a squirmer in the presence of an infinite lattice of inert spherical obstacles and identified the regimes based on the obstacle packing fraction. Bhattacharjee et al. \cite{Bhattacharjee2019} experimentally demonstrated the hopping and trapping behavior of bacteria in the presence of a confined heterogeneous environment. Ortlieb et al. \cite{Ortlieb2019} presented the distribution of the colloidal displacements in the presence of the \textit{Chlamydomonas reinhardtii}. To the best of our knowledge, although microswimmer dynamics in external flows have been widely studied, studies on the translational and rotational behavior of microswimmers in a channel flow with the presence of passive colloids remain limited.

In the present study, we focus on the effect of the colloid-laden channel flow on the motion of a microswimmer. We implement the background fluid and the microswimmer using the MPCD model \cite{Malevanets1999, Malevanets2000, Tao2008, Gotze2010, Mandal2021, Ramprasad2025} and the squirmer model \cite{Lighthill1952, Blake1971, Gotze2010}, respectively. This approach ensures accurate capture of hydrodynamic interactions while automatically accounting for thermal fluctuations. The colloids are considered as hard spherical objects. Here, the squirmer is characterized as a pusher ($\beta < 0$), neutral ($\beta = 0$), or a puller ($\beta > 0$) type microswimmer, where $\beta$ is the squirmer parameter. We explicitly validate the implementation of the MPCD and squirmer model, the colloidal suspension, sphere near-wall interactions, and the squirmer in a channel flow. We investigate the combined effects of flow and colloidal confinement on squirmer dynamics between parallel walls. We observe that the presence of colloids in this system causes the pusher to move towards the channel center, and the puller to move near the wall, with a dominant upstream orientation.

\section{\label{sec:method}Problem definition and methodology}

\begin{figure}[b]
    \centering
    \includegraphics[]{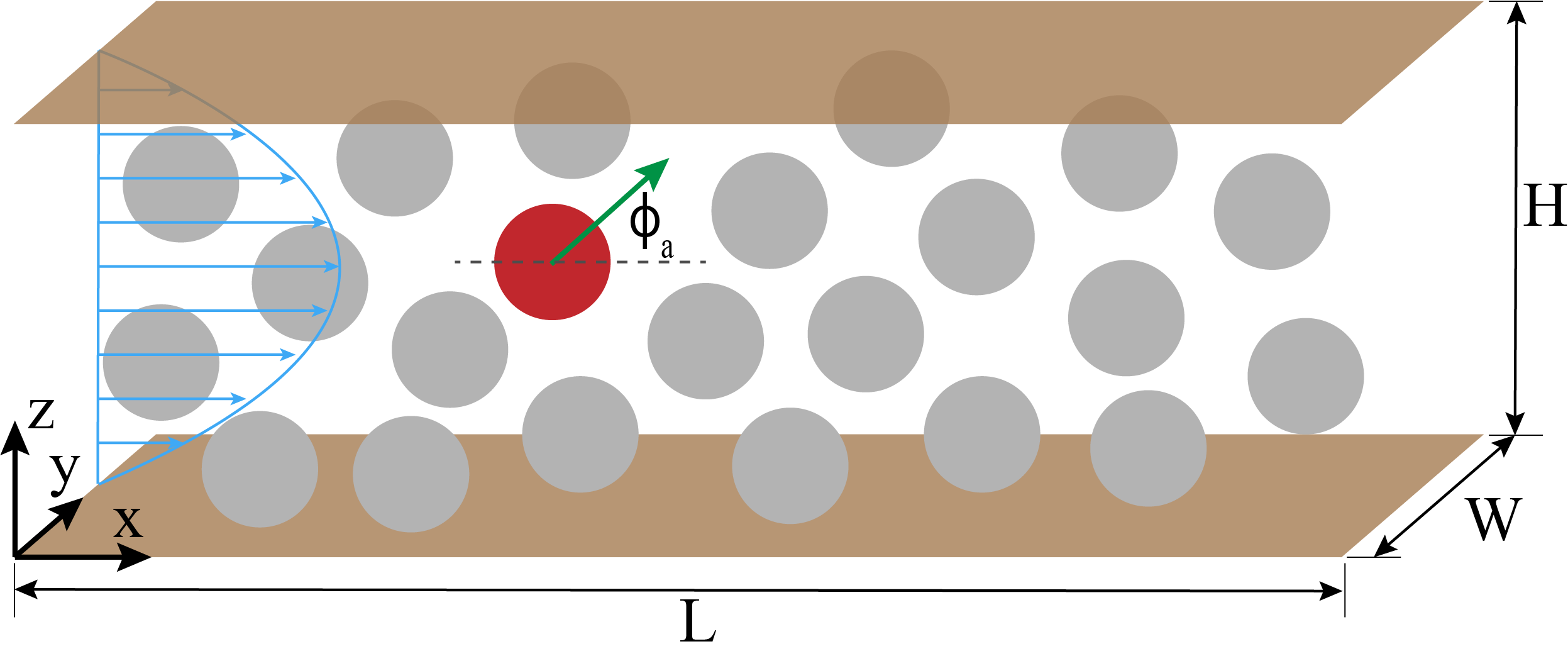}
    \caption{Schematic representation of simulation domain with a single squirmer (red), colloids (gray), and walls (top and bottom). The green arrow represents the squirmer orientation vector making an angle $\phi_a$ (azimuthal angle) with the flow direction (positive $x$-axis). Blue colored parabolic curve represents the fluid velocity profile along the positive $x$ direction.}
    \label{fig:prob_domain}
\end{figure}

To simulate the microswimmer in colloidal suspension between two walls (Fig. \ref{fig:prob_domain}), we adopted a mesoscale simulation framework called the multi-particle collision dynamics coupled with molecular dynamics (hybrid MPCD-MD) model \cite{Gotze2010, Ramprasad2025} to simulate the background fluid along with the squirmer \cite{Downton2009} to represent the microswimmer. We show the schematic representation of the simulation domain in Fig. \ref{fig:prob_domain} with size ($L, W, H$) in ($x, y, z$) directions, respectively. The top and bottom planes represent the walls at a distance $H$. Here, the squirmer (red sphere) is placed in the colloids (gray spheres), which are initially positioned using a uniform distribution. $\phi_a$ represents the azimuthal angle: the angle between the squirmer orientation vector (green arrow) and the flow direction (positive $x$ direction).

\subsection{\label{sec:num_model}Numerical model}

MPCD is a mesoscale model initially introduced by Malevanets and Kapral \cite{Malevanets1999}, which is a combination of streaming and collision steps. In this model, initially, $N$ number of particles of mass $m$, position $\bm{r}_i$, and velocity $\bm{v}_i$ are positioned in a three-dimensional domain with uniform distribution, considering $N_c$ number of particles per cell of cell size $a_0$. Initially the velocities of these particles are assigned using a Gaussian distribution with zero mean and unit variance ($k_BT = 1$). Here, $k_B$ is the Boltzmann constant and $T$ is the temperature. In the streaming step, the $N$ number of particles' positions ($\bm{r}_i$) and velocities ($\bm{v}_i$) are updated for a timestep of $\delta t$ (Eq. \ref{eqn:streaming}), where an equal and constant force $\boldsymbol{g}$ is applied on each MPCD particle in the positive $x$ direction \cite{Zottl2018, Qi2020rheotaxis} to implement the channel flow which acts as a plane Poiseuille flow, following which the particles are separated into cells of unit size ($a_0 = 1$). Here, $\bm{v}_i(t)$ represents the velocity of the $i^{th}$ particle at time $t$. The $\boldsymbol{g}$ is calculated from \cite{Qi2020rheotaxis} $\boldsymbol{g} = \frac{8mu_m\nu}{H^2} \boldsymbol{e}_x$, where $\nu$ is the kinematic viscosity of the MPCD fluid and $u_m$ is the centerline velocity of the fluid between two parallel plates with a theoretical velocity profile of $U_x/u_m = 4z(H-z)/H^2$. In the collision step, the momentum is conserved among the cells using Eq. \ref{eqn:collision}, which considers the Anderson thermostat (AT) and angular momentum conservation (+a) known as MPCD-AT+a collision rule \cite{Gotze2010}. Here, superscript $ran$ represents the random velocity generated from the Gaussian distribution, and subscripts $i$ and $j$ represent the particle count in the simulation domain and a particular cell $c$, respectively. $\bm{\Pi}_c$ is the moment-of-inertia tensor of the particles contained by cell $c$. To account for the Galilean invariance, we randomly shift the particles before and after the collision step.

\begin{subequations}\label{eqn:streaming}
\begin{eqnarray}\label{xstreaming}
\bm{r}_i(t+\delta t) = \bm{r}_i(t) + \bm{v}_i(t) \delta t + \bm{g}\frac{\delta t^2}{2m}.
\\\label{vstreaming}
\bm{v}'_i(t+\delta t) = \bm{v}_i(t) + \bm{g}\frac{\delta t}{m}
\end{eqnarray}
\end{subequations}

\begin{equation}\label{eqn:collision}
\begin{split}
\bm{v}_i(t+\delta t) & = \sum_{j\in cell} \bm{v}'_j(t+\delta t)/N_c + \bm{v}_i^{ran}(t) - \sum_{j\in cell} \bm{v}_j^{ran}(t)/N_c\\
& + m\bm{\Pi}_c^{-1}(t)\sum_{j\in cell}{\bm{r}_{j,c}(t)\times(\bm{v}'_j(t+\delta t)-\bm{v}_j^{ran}(t))}.
\end{split}
\end{equation}

 The squirmer model is a multimodal expansion of surface velocities of microorganisms, considered as a rigid sphere with radius $R_s$. This model was initially introduced by Lighthill \cite{Lighthill1952} and then further developed by Blake \cite{Blake1971}. The surface velocity field $\bm{v}_{sq}$ of the squirmer is expressed as a function of position on the squirmer surface relative to the squirmer centroid ($\bm{r}_{s}$) and orientation ($\bm{e}$) as follows  \cite{Zottl2018}: 
 
\begin{equation}\label{eqn:vsq_surf}
\bm{v}_{sq}(\bm{r}_{s}, \bm{e}) = B_1[1+\beta(\bm{e} \cdot \hat{\bm{r}}_{s})][(\bm{e} \cdot \hat{\bm{r}}_{s})\hat{\bm{r}}_{s}-\bm{e}],
\end{equation}
 
\noindent where $B_1$ is the self-propulsion strength and $\beta$ is the squirmer parameter ($\beta = B_2/B_1$). Here, $\hat{\bm{r}}_{s}$ is the unit vector from the squirmer center to a point on the surface where the MPCD particle is in contact with the squirmer. The theoretical swimming speed of the squirmer in a Newtonian fluid ($V_0$) is defined as ($2B_1/3$). Neglecting the higher modes, we considered the source dipole, force dipole, and source quadrupole to represent pusher ($\beta < 0$), neutral ($\beta = 0$), and puller ($\beta > 0$) types of microswimmers.

The colloids in the present system are modeled as rigid spheres of radius $R_c$, which are initially positioned in the domain using a uniform distribution. The colloidal packing fraction ($\phi$) is defined as $\phi = 4\pi R_c^3 N_{coll}/(3LWH)$. Here, $N_{coll}$ is the number of colloids. To implement two parallel plates, we considered walls of unit cell size thickness in the $z$ direction at a distance of $H$. The collision of a fluid particle with the squirmer, colloid, and wall is implemented with a half-time bounce-back rule. Here, the velocity of the $i^{th}$ fluid particle, which is in collision with the wall, is updated by $\bm{v}'_i = -\bm{v}_i$. The particle in collision with the squirmer is updated by $\bm{v}'_i = -\bm{v}_i+2\bm{v}_s$, where $\bm{v}_s$ indicates absolute surface velocity of the squirmer at the point of collision \cite{Ramprasad2025}, which includes the surface slip velocity of the squirmer ($\bm{v}_{sq}$) (Eq. \ref{eqn:vsq_surf}). Here, $\bm{v}_{s}(\bm{r}_{s},\bm{e})= \bm{V} + \bm{\Omega}\times(\bm{r}_{s}-\bm{R}) + \bm{v}_{sq}(\bm{r}_{s}, \bm{e})$. The translational ($\bm{V}$) and angular ($\bm{\Omega}$) velocity of the squirmer are updated to conserve the linear and angular momentum with the sum of the change in momentum of the colliding MPCD particles with the squirmer \cite{Zottl2018, Ramprasad2025}. The particle in collision with a colloid is updated by $\bm{v}'_i = -\bm{v}_i+2\bm{v}_c$, where $\bm{v}_c$ is the absolute colloid velocity without the surface slip velocity. Similar to the squirmer, the translational velocity of the colloids is updated with the sum of the change in momentum of the colliding MPCD particles with the colloids. The squirmer-colloid, squirmer-wall, colloid-colloid, and colloid-wall interactions are implemented with the cut-off Weeks-Chandler-Anderson (WCA) potential \cite{Gompper2009, Zottl2019, Martin2025, Ramprasad2025} as follows:

\begin{eqnarray}\label{eqn:WCA}
U (r) = \biggl\{
\begin{array}{c}
4\epsilon \left[ \left( \frac{\sigma}{r} \right)^{12}-\left(\frac{\sigma}{r} \right)^6 \right]+\epsilon, \; \; r<2^{1/6}\sigma\\
0, \; \; \; \; \; \; \; \; \; \; \; \; \; \; \; \; \; \; \; \; \; \; \; \; \; \; \; \; \; \; \; \; \; \; r\geq2^{1/6}\sigma
\end{array}
\end{eqnarray}

\noindent where $\epsilon$ is the potential strength, $\sigma$ is the sum of the squirmer and colloid radius ($R_s + R_c$) in case of squirmer colloid interaction, colloid diameter ($2R_c$) in case of colloid-colloid interaction, and radius of squirmer or colloid in case of the squirmer-wall or colloid-wall interaction. The interaction force is calculated from the derivation of the potential from Eq. \ref{eqn:WCA} as $\bm{F}(t) = -\bm{\nabla} U(r)$. The position ($\bm{R}$), orientation ($\bm{e}$), and velocity ($\bm{V}$) of the squirmer with mass $M_s$ is updated using the velocity Verlet algorithm \cite{Ramprasad2025} with a timestep of $\delta t_s = \delta t/10$ as shown in Eq. \ref{eqn:svva}. Similarly, the position and velocity of the colloids are updated, whereas the orientation of the colloids is not considered.

\begin{subequations}\label{eqn:svva}
\begin{eqnarray}
\bm{R}(t+\delta t_s) = \bm{R}(t) + \bm{V}(t)\delta t_s + \bm{F}(t)\frac{\delta t_s^2}{2M_s}, \label{vvaa}
\\
\bm{e}(t+\delta t_s) = \bm{e}(t) + (\bm{\Omega}(t)\times\bm{e}(t))\delta t_s,
\\
\bm{V}(t+\delta t_s/2) = \bm{V}(t) + \bm{F}(t)\frac{\delta t_s}{2M_s}, \label{vvab}
\\
\bm{V}(t+\delta t_s) = \bm{V}(t+\delta t_s/2) + \bm{F}(t+\delta t_s)\frac{\delta t_s}{2M_s}. \label{vvac}
\end{eqnarray}
\end{subequations}

\noindent In the present study, we characterize the flow by a non-dimensional parameter, flow strength ($V_r$), defined as $V_r = u_m/V_0$. As the MPCD and the squirmer model have great advantages in parallelization, we have developed an in-house CUDA-C program for the GPU to simulate the current model.

\subsection{Validations}

The validations for the MPCD code and the squirmer model are detailed in our previous work \cite{Ramprasad2025}. Due to the unavailability of comparable works in the published literature, we individually compared and validated the current model implementation for colloidal suspensions and a squirmer in a plane Poiseuille flow. To verify the implementation of colloidal suspension and the hydrodynamic interactions, we simulated a suspension of spherical colloids in a periodic domain \cite{Winkler2005} and validated the effective diffusion coefficient ($D_{eff}$) with respect to the colloidal packing fraction ($\phi$) and confirmed the Gaussianity of colloidal displacement distribution (see Appendix \ref{sec:valid_coll}). To verify the hydrodynamic interactions between a sphere and the wall, we studied the force and torque coefficients of a sphere with effective torque or in the presence of Poiseuille flow (see Appendix \ref{sec:valid_sphere}). To verify the implementation of squirmer in plane Poiseuille flow, we compared the translational and directional probabilities of squirmer with Qi et al. \cite{Qi2020rheotaxis} (see Appendix \ref{sec:valid_qi}).

\subsection{Model parameters}
\label{sec:para}

In the present work, we consider neutrally buoyant squirmer and colloids of radius $R_s = R_c = 4a_0$. The simulation domain consists of two walls at a distance $H = 10R_s$ in the $z$ direction (velocity gradient direction) as shown in Fig. \ref{fig:prob_domain}. The length of the domain along the $x$ direction (flow direction) is $L = 30R_s$, and the width along the $y$-axis (vorticity direction) is $W = 20R_s$. We consider the cell size ($a_0$) as unity, and the number of particles per cell ($N_c$) as 30. The MPCD timestep ($\delta t$) is $0.02t_0$, and the velocity-Verlet timestep ($\delta t_s$) is $0.002t_0$. The cut-off WCA potential strength ($\epsilon$) is considered as 1000$k_B T$ to implement the hard spherical squirmer and colloids, and to avoid the overlap. With the above parameters, we calculated the squirmer theoretical speed ($V_0 = 2B_1/3$), the Reynolds number ($Re = V_0R_s/\nu$), and the Péclet number ($Pe = 2V_0R_s/D$, with $D = k_BT/(6\pi\nu N_cR_s)$ as the translational diffusion coefficient) as shown in the Table \ref{tab:MPCDparameters}. We choose the parameters in such a way that $Re<1$, to ensure a low Reynolds number environment.

\begin{table}[h]
    \centering
    \caption{\label{tab:MPCDparameters} Considered MPCD parameters.}
        \begin{tabular}{ccccc}
            \hline
             S. No.&$B_1$&$V_0$& $Re$ & $Pe$
            \\ \hline
            1&	0.02076&	0.0138& 0.028&	250\\
            2&	0.04155&	0.0277& 0.056&	500\\
            3&	0.06225&	0.0415& 0.083&	750\\
            \hline
        \end{tabular}
\end{table}

For the channel flow case, we limit the study to $Pe$ = 750, where the influence of thermal fluctuations is minimal compared to the strong advective flow. To study the effect of flow on the squirmer motion, we consider flow strengths ($V_r$) of 0.5, 2, and 4. To check the compressibility effects, we calculated the Mach number ($Ma = v_{max}/\sqrt{5k_BT/(3m)}$) \cite{Nikoubashman2014flow}, where $v_{max}$ is the maximum speed of the squirmer ($V_0$) or the flow ($u_m$). For our highest considered $u_m$ value, 0.166, we get $Ma$ = 0.129, which is less than the threshold limit 0.25, above which the compressibility effects become significant \cite{Nikoubashman2014flow, Lamura2001Multi-particleCylinder}.

\section{\label{sec:R_D}Results and Discussion}

\subsection{Squirmer in colloid-laden channel flow}
\label{ssec:with_flow} 

In the present study, we assume that the squirmer is initially ($t = 0$) oriented in the downstream direction ($\phi_a<\pi/2$ or $\phi_a>3\pi/2$) at different locations along the velocity gradient direction to simplify the study conditions. Each data set is an average or a sum of 5 realizations, each $2-4\times10^6$ timesteps.

\begin{figure}[h]
    \centering
    \includegraphics[width = 0.9\linewidth]{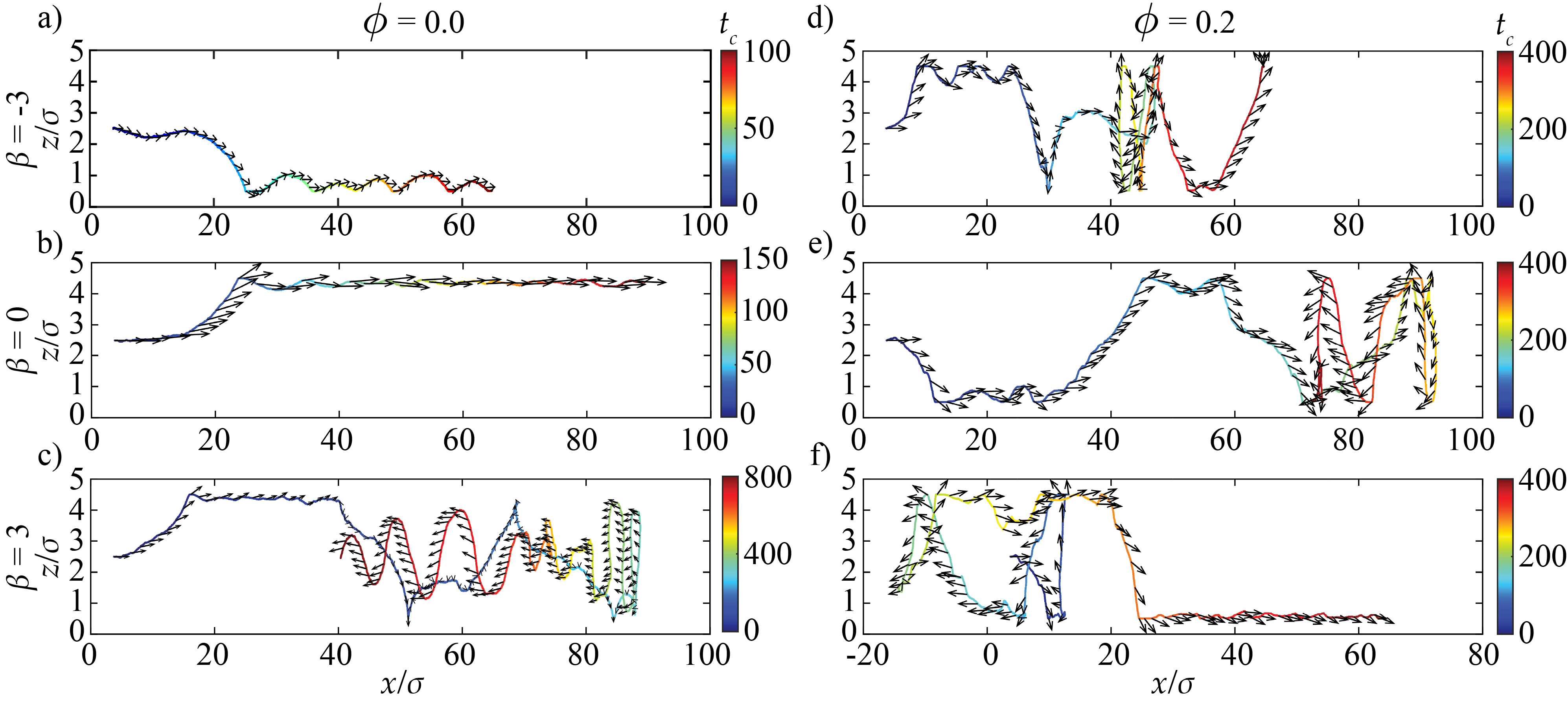}
    \caption{Trajectory of different types of squirmers along with the instantaneous orientations (arrows) for (a, d) pusher ($\beta = -3$), (b, e) neutral ($\beta = 0$), and (c, f) puller ($\beta = 3$) for (a-c) $\phi = 0$ and (d-f) $\phi = 0.2$ with $V_r = 0.5$. Here, the colorbar indicates the characteristic time of the squirmer.}
    \label{fig:traj_Pe750_allbeta_phi00_20_ub05_400tc_ds}
\end{figure}

\subsubsection{Translational and directional behaviors of the squirmer}
\label{ssec:traj}

Fig. \ref{fig:traj_Pe750_allbeta_phi00_20_ub05_400tc_ds} depicts the trajectory of the squirmer with instantaneous orientations of the squirmer to illustrate the upstream and downstream directionality of the squirmer. Here, $t_c$ represents the characteristic time, defined as the time taken by the squirmer to travel its radius $R_s$ with theoretical velocity $V_0$. Without colloids and at low flow strengths ($V_r = 0.5$), the pushers and neutral squirmers follow the wall with downstream orientation (Fig. \ref{fig:traj_Pe750_allbeta_phi00_20_ub05_400tc_ds}(a, b)), whereas pullers swim against the flow, migrating towards the channel center (Fig. \ref{fig:traj_Pe750_allbeta_phi00_20_ub05_400tc_ds}c). At high flow strengths  ($V_r = 4$), all squirmers show oscillatory cross-stream trajectories between two walls (see Appendix \ref{assec:traj_Vr4}). Such behaviors are observed in experiments \cite{Hill2007experiment, Barry2015chlamydomonasrheo, Omori2022chlaydomonasrheo} and modeled in numerical studies \cite{Zottl2012, Qi2020rheotaxis}. However, heterogeneity in the system, such as the colloidal suspension in our present study, affects this behavior. The heterogeneity in this system comes from the non-uniform distribution of the colloids influenced by the squirmer interactions. With the presence of colloids, at high packing fraction ($\phi = 0.2$) and at low flow strength ($V_r = 0.5$), the pusher and neutral-type squirmers have dominant upstream orientation in the long time range (Fig. \ref{fig:traj_Pe750_allbeta_phi00_20_ub05_400tc_ds}(d, e)), whereas the pullers are following the wall with a dominated downstream orientation in the long time range (Fig. \ref{fig:traj_Pe750_allbeta_phi00_20_ub05_400tc_ds}f), in contrast to the case without colloids (Fig. \ref{fig:traj_Pe750_allbeta_phi00_20_ub05_400tc_ds} (a-c)). The continuous interactions of the pusher and neutral squirmers with the colloids in the system cause it to reorient in the hydrodynamically stable upstream direction \cite{Zottl2012}. In the case of the pullers, the dominant hydrodynamic repulsion forces from the colloidal suspension cause them to move away from the channel center \cite{Spagnolie2015, Alonso-Matilla2019}. The translational and rotational motion of the squirmer is subjected to the stochasticity in the system. To systematically characterize this behavior, we performed multiple realizations and studied a joint probability distribution of position and orientation.

\begin{figure}
    \centering
    \includegraphics[width = \linewidth]{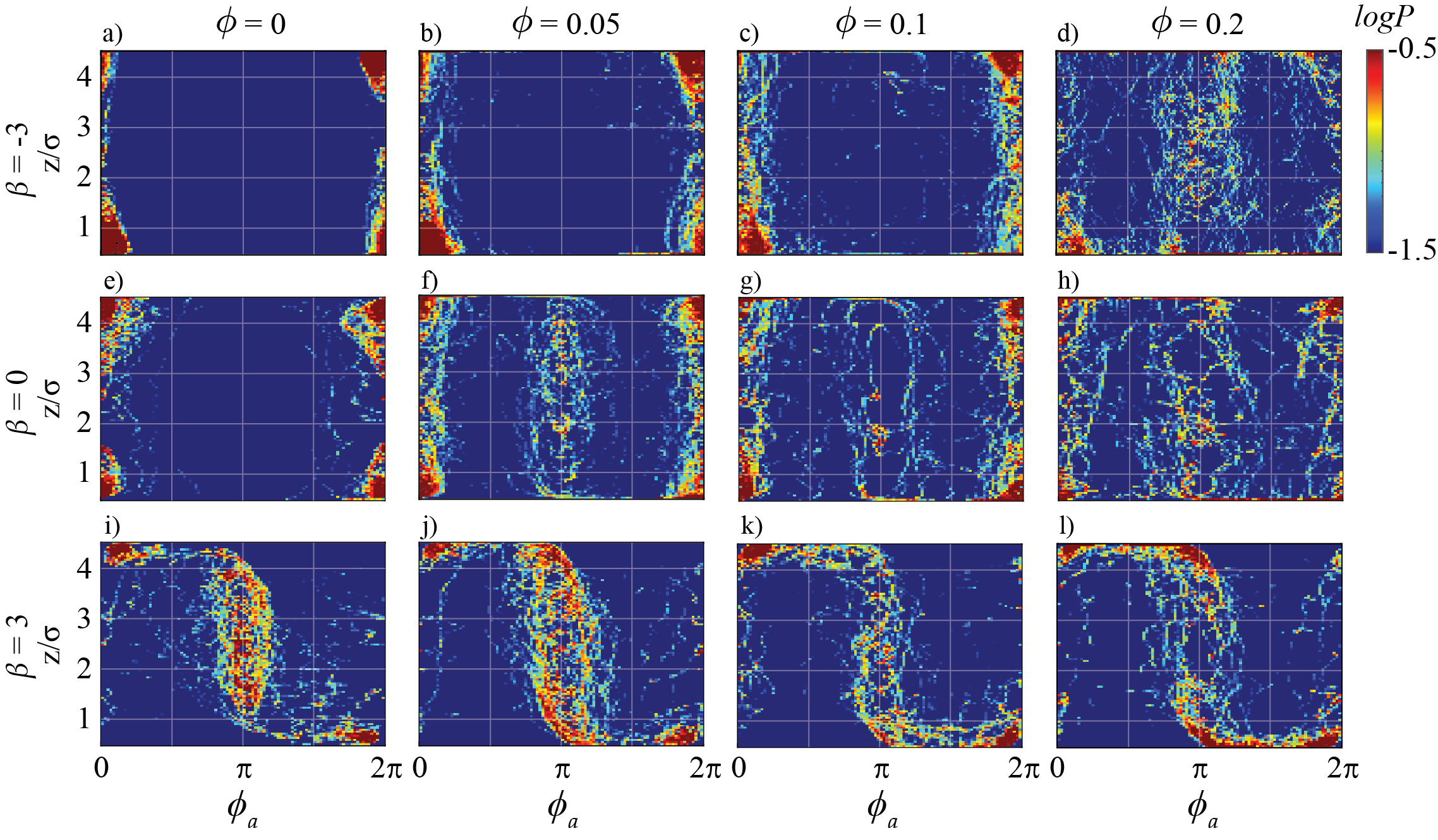}
    \caption{Joint probability distribution function ($P(z,\phi_a)$) of squirmer position along velocity gradient direction (z direction) and azimuthal angle ($\phi_a$) with the flow strength ($V_r = 0.5$) for different types of squirmers: (a-d) pusher ($\beta = -3$), (e-h) neutral ($\beta = 0$), and (i-l) puller ($\beta = 3$), and colloidal packing fraction ($\phi = 0, 0.05, 0.1,\text{and} 0.2$). Here, the color bar shows the logarithmic probability distribution.}
    \label{fig:PDF_Pe750_allbeta_allphi_ubU05}
\end{figure}

\begin{figure}
    \centering
    \includegraphics[width = \linewidth]{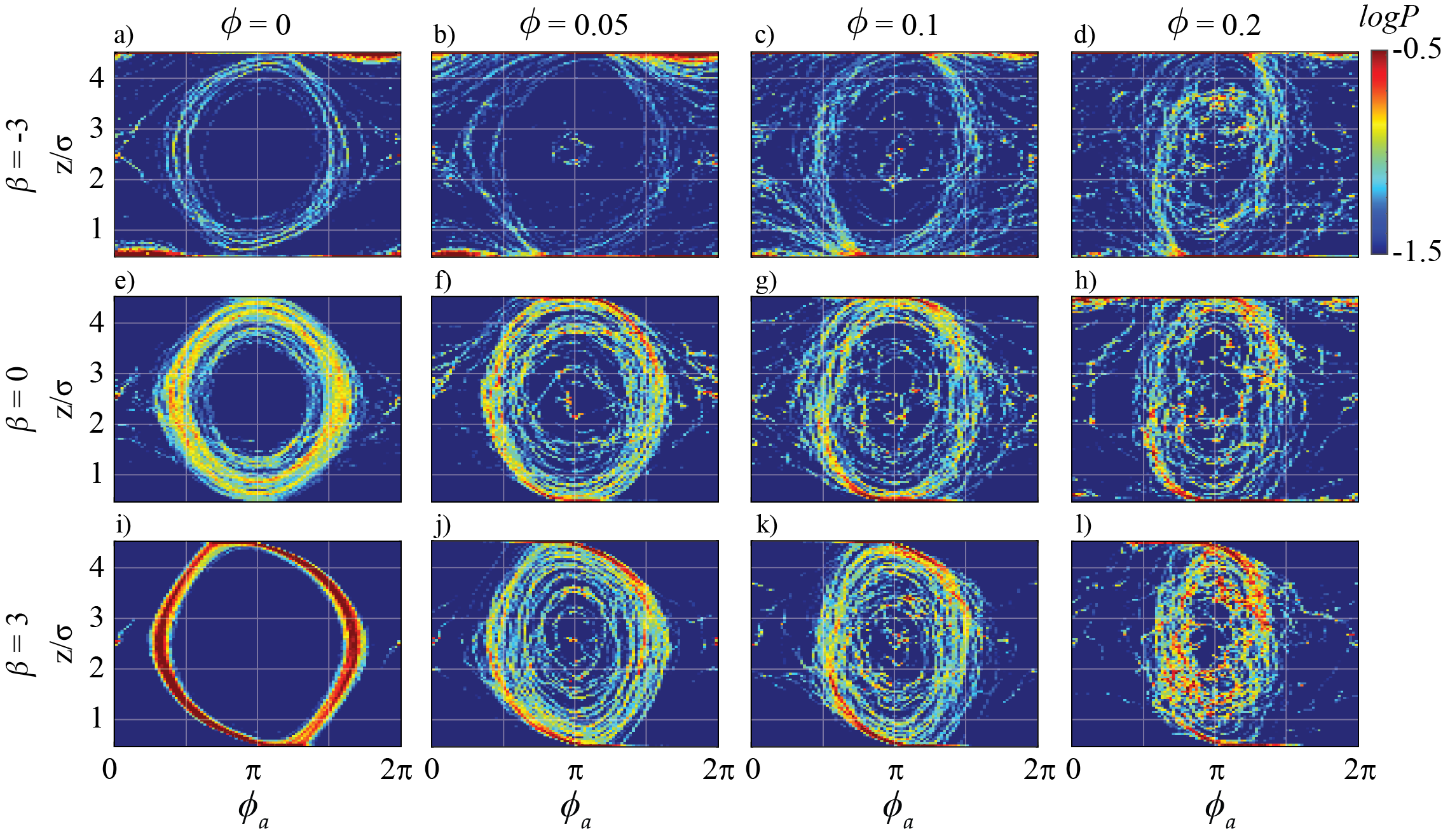}
    \caption{Joint probability distribution function ($P(z,\phi_a)$) of squirmer position along velocity gradient direction (z direction) and azimuthal angle ($\phi_a$) with the flow strength ($V_r = 4$) for different types of squirmers: (a-d) pusher ($\beta = -3$), (e-h) neutral ($\beta = 0$), and (i-l) puller ($\beta = 3$), and colloidal packing fraction ($\phi = 0, 0.05, 0.1, \text{ and } 0.2$). Here, the color bar shows the logarithmic probability distribution.}
    \label{fig:PDF_Pe750_allbeta_allphi_ubU40}
\end{figure}

\subsubsection{Probability distribution of the squirmer position and the orientation}
\label{ssec:jpdf}

To study the probabilistic behavior of position and orientation of the squirmer, we calculated the joint probability distribution function ($P(z,\phi_a)$) with respect to the squirmer position along the velocity gradient direction ($z$) and the azimuthal angle ($\phi_a$) as shown in the Figs. (\ref{fig:PDF_Pe750_allbeta_allphi_ubU05}, \ref{fig:PDF_Pe750_allbeta_allphi_ubU40}) for $V_r = 0.5$ and $4$, respectively. In both the figures, we show the logarithmic probability contours for different squirmer types: $\beta = -3, 0, \text{ and } 3$, for colloidal packing fractions ($\phi$) $0, 0.05, 0.1, \text{ and } 0.2$. Here, $\phi = 0$ indicates the case without colloids. From Fig. \ref{fig:PDF_Pe750_allbeta_allphi_ubU05}(a, e, i), we observe that, at flow strength ($V_r = 0.5$), the pushers are preferentially accumulated near the wall with downstream orientation, whereas the pullers are oriented in the upstream direction near the center of the channel. We also observe a higher positional probability near the walls for the downstream-oriented squirmers. Such behaviors are qualitatively consistent with experimental observations of shear-induced alignment and rheotactic behavior in microswimmers \cite{Hill2007experiment, Rusconi2014, Barry2015chlamydomonasrheo}. From Fig. \ref{fig:PDF_Pe750_allbeta_allphi_ubU40}(a, e, i), we observe that, at high flow strength ($V_r = 4$), the pushers are following the wall with downstream direction, and pullers are more probable at the wall with upstream direction and near the center of the channel with downstream direction. Here, the ring-like contours represent the squirmer oscillating between the walls. We observe that at low flow strength ($V_r = 0.5$), the pushers are attracted towards the center of the channel with the dominated upstream orientation ($\pi/2 < \phi_a < 3\pi/2$) (Fig. \ref{fig:PDF_Pe750_allbeta_allphi_ubU05}(a-d)) as the $\phi$ increases. This is due to the increasing effective hydrodynamic attraction between colloids and the pusher \cite{Gotze2010, Molina2014}, which dominates over the wall-squirmer hydrodynamic attraction \cite{Llopis2010} and guides the pusher towards a higher symmetric region. In the case of a neutral-type squirmer, the absence of a dominant force-dipole component leads to weak hydrodynamic interactions, which result in less structured and more diffusive motion in the channel (Fig. \ref{fig:PDF_Pe750_allbeta_allphi_ubU05}(e-h)). Contrary to the pusher translational behavior, at low flow strength ($V_r = 0.5$), the pullers are more probable to move along the wall, that is, swimming away from the colloids with dominating upstream orientation (Fig. \ref{fig:PDF_Pe750_allbeta_allphi_ubU05}(i-l)). This is due to the contractile flow field of the pullers that increases the effective hydrodynamic repulsion between the colloids and the puller \cite{Molina2014}, and the colloids being trapped in the center of the channel, which is the shear-induced trap region \cite{Rusconi2014, Qi2020rheotaxis}. 

At low flow strength with the presence of colloids, the squirmer-induced hydrodynamic interactions dominate the imposed background flow. Consequently, pushers and pullers, characterized by extensile and contractile stresslet fields, exert significant effective hydrodynamic attraction towards and repulsion from the colloids \cite{Gotze2010, Llopis2010}. However, with increased flow strength, the effect of the squirmer-induced hydrodynamic interactions will be less significant than that of the background channel flow. Consequently, advection primarily governs the translation dynamics, while shear-induced torques dominate the bulk orientational dynamics, with hydrodynamic interactions remaining important near walls and colloids \cite{Zottl2012, Qi2020rheotaxis}. At high flow strength ($V_r = 4$), in the absence of colloids, cross-streaming of the squirmers (squirmer facing the wall) is observed (Fig. \ref{fig:PDF_Pe750_allbeta_allphi_ubU40}(a, e, i)), consistent with previous studies of microswimmers in Poiseuille flow \cite{Zottl2012}. This decreases as the colloidal packing fraction increases due to enhanced scattering and hindrance from the colloids, as shown in the Fig. \ref{fig:PDF_Pe750_allbeta_allphi_ubU40}. Notably, for pushers at high flow strength ($V_r = 4$), we observe the increase in the probability of upstream orientation near the walls, similar to the low flow strength ($V_r = 0.5$) case, due to the wall-induced hydrodynamic interactions that remain significant even when the background flow dominates.

\subsubsection{Alignment of the squirmer with flow direction}
\label{ssec:orientation}

\begin{figure}[t]
    \centering
    \includegraphics[width = 0.8\linewidth]{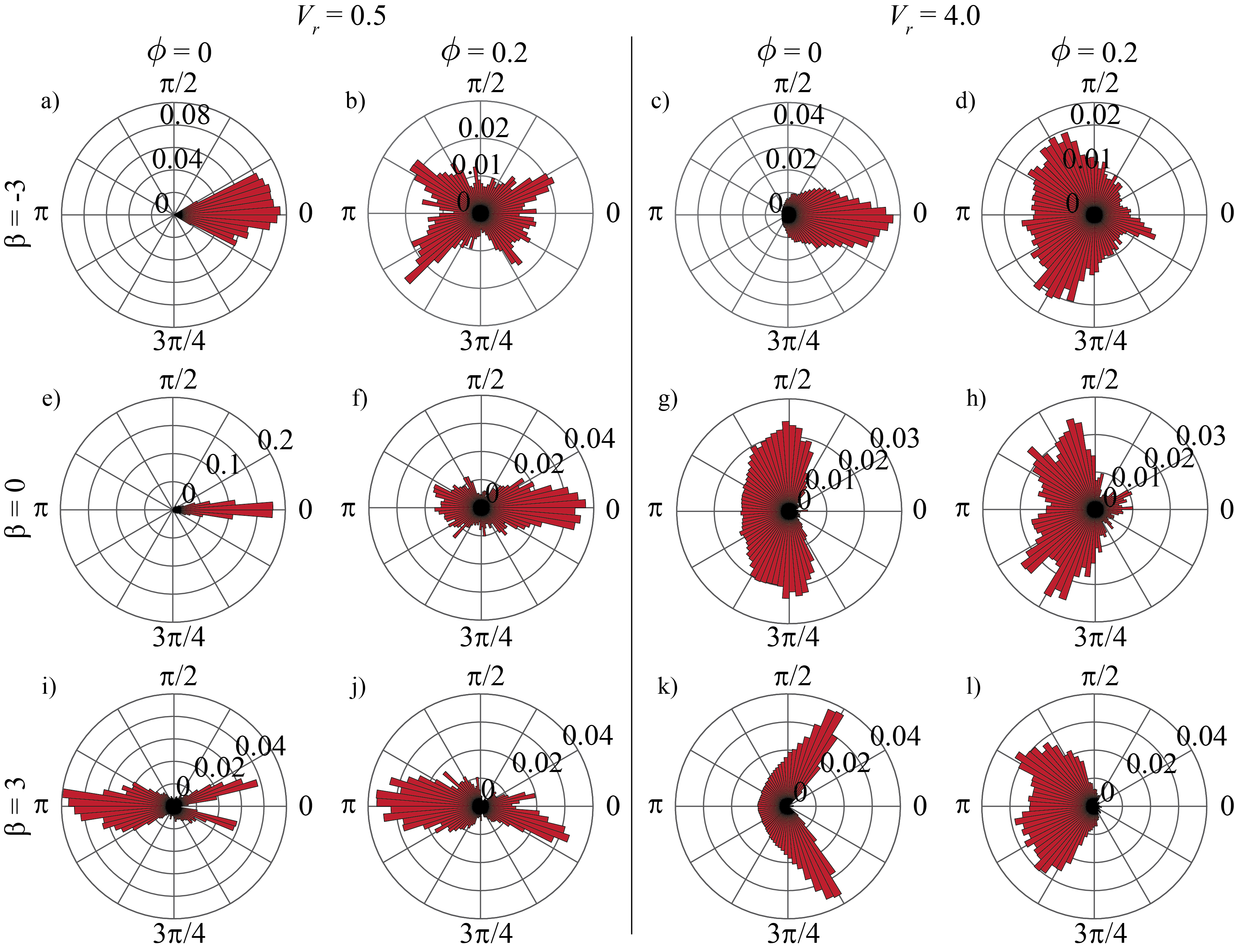}
    \caption{Orientation distribution of (a-d) pusher ($\beta = -3$), (e-h) neutral ($\beta = 0)$, and (e-h) puller ($\beta = 3$) for (a, b, e, f, i, j) $V_r = 0.5$ and (c, d, g, h, k, l) $V_r = 4.0$. Each case is compared without ($\phi = 0$) and with ($\phi = 0.2$) colloids.}
    \label{fig:phi_a_pdf_Pe750_ball}
\end{figure}

The squirmer orientation with respect to the flow direction ($\phi_a$) is essential to assess the behavior of different types of squirmers. The orientation distribution of the squirmer is shown in Fig. \ref{fig:phi_a_pdf_Pe750_ball}. It is evident from Fig. \ref{fig:phi_a_pdf_Pe750_ball} that, for $\phi = 0$, the increased flow strength will increase the cross-stream orientation of the squirmer, which depicts the oscillatory motion of the squirmer between two walls, which is observed in previous studies \cite{Zottl2012, Qi2020rheotaxis, Dey2022OscillatoryMicrochannels}. Supporting the observations in Figs. (\ref{fig:PDF_Pe750_allbeta_allphi_ubU05}, \ref{fig:PDF_Pe750_allbeta_allphi_ubU40}), we observe that the increase in $\phi$ decreases the probability of the downstream orientation ($\phi_a < \pi/2$ or $\phi_a > 3\pi/2$) and increases the probability of the upstream orientation ($\pi/2<\phi_a<3\pi/2$) of the pusher for all $V_r$ (0.5 to 4) (Fig. \ref{fig:phi_a_pdf_Pe750_ball}(a-d)). In the case of a neutral squirmer, at low flow strength ($V_r = 0.5$), we observe a conversion of downstream-oriented motion to upstream-oriented motion as the $\phi$ increases. Whereas at high flow strength ($V_r = 4$), the cross-stream oriented motion is converted to upstream oriented motion (Fig. \ref{fig:phi_a_pdf_Pe750_ball}(e-h)). For pullers, the changes in upstream and downstream orientations are insignificant for low flow strength ($V_r = 0.5$) (Fig. \ref{fig:phi_a_pdf_Pe750_ball}(i, j)). However, at high flow strength ($V_r = 4.0$), we observe an increase in upstream orientation and a decrease in cross-stream orientation (Fig. \ref{fig:phi_a_pdf_Pe750_ball}(k, l)).

\subsubsection{Streamwise velocity of the squirmer}
\label{ssec:velocity}

\begin{figure}[h]
    \centering
    \includegraphics[width = 0.7\linewidth]{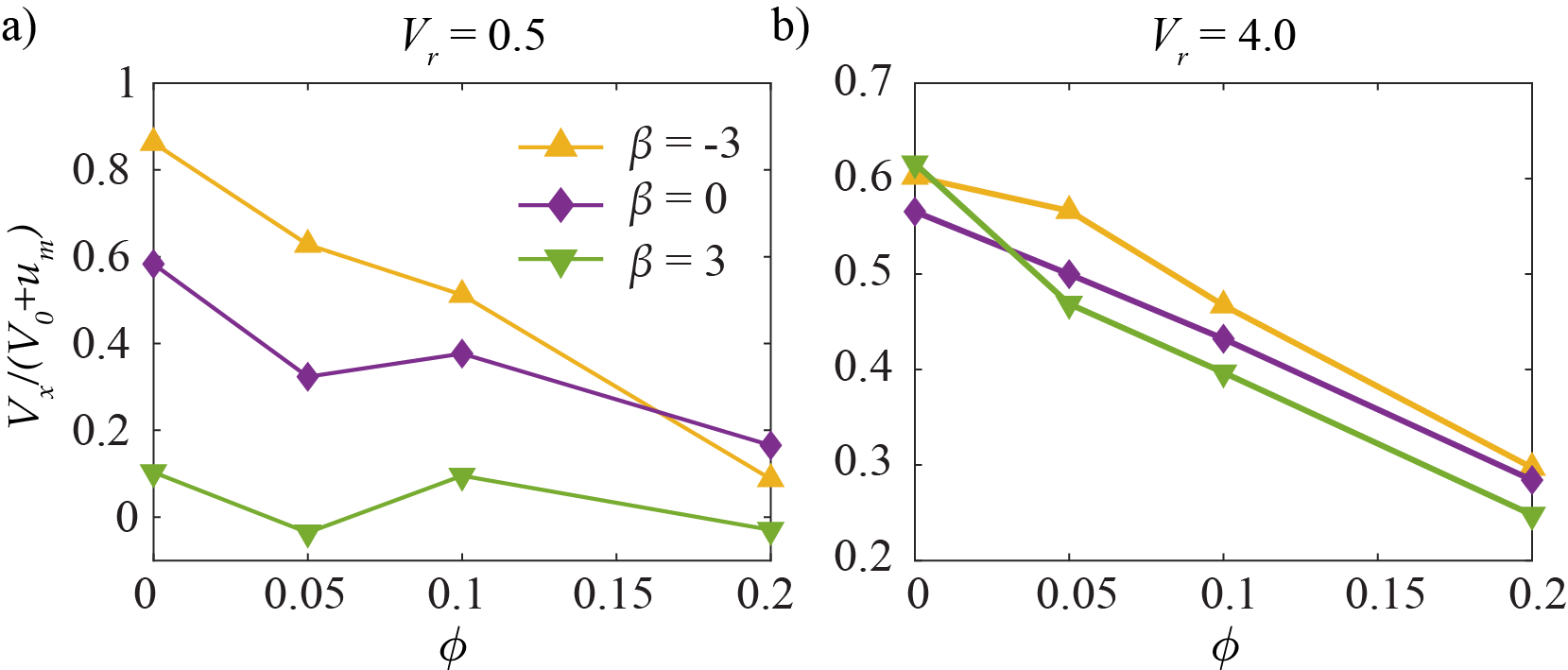}
    \caption{Mean streamwise velocity ($V_x$) of different types of the squirmer in the flow direction with respect to the colloidal packing fraction ($\phi$) for flow strengths a) $V_r = 0.5$ and b) $V_r = 4$. Here, the mean streamwise velocity of the squirmer is normalized by the sum of the squirmer's theoretical speed ($V_0$) and the maximum flow speed ($u_m$).}
    \label{fig:v_mean_Pe750_ball_ubU05_4_vx}
\end{figure}

The increase in $\phi$ reduces the motion of the squirmer in the flow direction due to enhanced collisions and hydrodynamic hindrance from the surrounding colloids. Predictably, this results in a reduction in the mean streamwise velocity of the squirmer ($V_x$) with increasing $\phi$ for all swimmer types and flow strengths ($V_r$) (Fig. \ref{fig:v_mean_Pe750_ball_ubU05_4_vx}). This reduction mainly reflects the suppression of effective transport arising from reduced free path and increased scattering in a crowded environment. Fig. \ref{fig:v_mean_Pe750_ball_ubU05_4_vx} illustrates the mean streamwise velocity ($V_x$), which is normalized with the sum of the theoretical squirmer speed ($V_0$) and the maximum flow speed ($u_m$) with respect to the colloidal packing fraction ($\phi$). The non-dimensional parameter $V_x/(V_0 + u_m)$ quantifies the effective downstream transport of the squirmer relative to the combined swimming and imposed-flow velocity scales. This helps us to compare different flow strengths, colloid concentrations, and squirmer types on the same scale.

At low flow strength ($V_r = 0.5$), the mean streamwise velocity shows a strong dependence on the squirmer type (Fig. \ref{fig:v_mean_Pe750_ball_ubU05_4_vx}a). We observe that the pushers ($\beta = -3$) have higher velocities compared to the pullers ($\beta = 3$), consistent with the rheotactic behavior reported in \cite{Qi2020rheotaxis}. These differences arise from their orientational dynamics: pushers preferentially align along the downstream direction and pullers in the upstream direction. Here, the pronounced difference in mean velocity between pushers and pullers at low flow strength indicates that the squirmer transport is primarily governed by squirmer-induced hydrodynamic interactions and activity-driven reorientation. In contrast, at high flow strength ($V_r = 4$), the background flow dominates the squirmer-induced hydrodynamic interactions \cite{Qi2020rheotaxis}. The mean streamwise velocity of all the squirmer types decreases with the increase in $\phi$, but shows a reduced sensitivity towards the squirmer parameter ($\beta$) (Fig. \ref{fig:v_mean_Pe750_ball_ubU05_4_vx}b). This indicates a transition from an activity-dominated regime at low $V_r$ to a flow-dominated regime at high $V_r$.  In a special case of $V_r = 0$ discussed in Appendix \ref{sec:without_flow}, the puller ($\beta = 3$) exhibits higher mean speed compared to the pusher ($\beta = -3$). This behavior is observed in pullers, who transport more effectively in the presence of obstacles, arising from their tendency to avoid prolonged interactions with colloids, whereas pushers experience stronger trapping \cite{Ramprasad2025}. Furthermore, in the absence of imposed flow, the shear-induced orientation changes are not present, which leads to the pure effects of squirmer-induced hydrodynamic interactions and stochastic effects.

To further elucidate these observations, we examine the spatial distribution of the time-averaged local streamwise velocity ($\overline{V}_x$), across the channel height, in the $z$-direction (see Appendix \ref{assec:velocity}). Due to the parabolic profile of the background Poiseuille flow, the squirmer velocity is inherently non-uniform along the velocity gradient direction \cite{Qi2020rheotaxis}. At low flow strength ($V_r = 0.5$), the velocity profiles showed a significant deviation from the parabolic shape, along with enhanced fluctuation indicating the dominant squirmer-induced hydrodynamic interactions and stochastic effects. However, at high flow strength ($V_r = 4$), the velocity profiles closely follow the imposed parabolic flow profile, indicating the dominant advection. Overall, these results highlight a transition from dominant hydrodynamic interaction at low flow strength to advection-dominated transport at high flow strength, with colloidal hindrance ($\phi$) primarily controlling the magnitude of transport, the mean streamwise velocity, and the squirmer parameter ($\beta$) modulating it through orientation-dependent effects.

\section{\label{sec:Conclusion}Conclusion}

The effects of thermal fluctuations, velocity gradient, and hydrodynamic interactions with the colloids are significant on the microswimmer's translational and directional motion. To numerically simulate the motion of the microswimmer in a colloid-laden channel flow, we considered multi-particle collision dynamics and the spherical squirmer model. An extensive validation of the in-house numerical code and model implementation has been carried out for passive colloidal suspensions, channel flow, squirmer, and near-wall hydrodynamic interactions.

We investigate a single squirmer in a colloidal suspension under a plane Poiseuille flow between two parallel plates. We observe that the presence of colloids significantly affects both the translational and orientational behavior of the squirmer. At low flow strength, pushers preferentially occupy the center of the channel, a more symmetric region, whereas pullers are biased towards the walls due to effective hydrodynamic interactions with the surrounding colloids. Additionally, at low flow strength, the pusher tends to adopt a dominant upstream orientation at higher colloidal packing fractions. At high flow strength, in the absence of colloids, both the pushers and pullers exhibit oscillatory motion between the top and bottom walls, accompanied by significant cross-stream orientation. In the presence of colloids, this cross-stream orientation is suppressed, and an increased upstream orientation is observed for all types of squirmers, due to the dominance of background flow and colloidal interactions over squirmer hydrodynamics and stochastic effects. From the velocity analysis, we observe that the mean streamwise velocity of the squirmer decreases with increasing colloidal packing fractions, with the extent of reduction depending on the squirmer type. In the presence of flow and colloids, pushers exhibit higher streamwise velocity than the pullers, in contrast to the case without flow. This difference is more pronounced at low flow strength than at high flow strength. The findings from the current work provide new insights into the coupled translational and rotational dynamics of microswimmers by systematically examining the interplay among swimmer-induced hydrodynamics, background flow, and suspended colloidal obstacles. The continuation of the present study can analyze the effects of the microswimmer's shape, size ratio, and polydispersity of the colloids, polymer suspension, and other relevant factors.

\section*{ACKNOWLEDGEMENTS}
 SM acknowledges the generous support of the Saroj Poddar Trust and the Anusandhan National Research Foundation (Grant No. EEQ/2021/000561). MR acknowledges the computing resources at HPCE, IIT Madras, and thanks Dr. Siddhant Mohapatra for helpful discussions. PSM acknowledges the V. Ganesan Faculty Fellowship received from IIT Madras.

\appendix

\section{\label{sec:valid}Validations}

\subsection{\label{sec:valid_coll}Validations: Colloidal suspension}

\begin{figure}[ht]
    \centering
    \begin{overpic}[width=\linewidth]{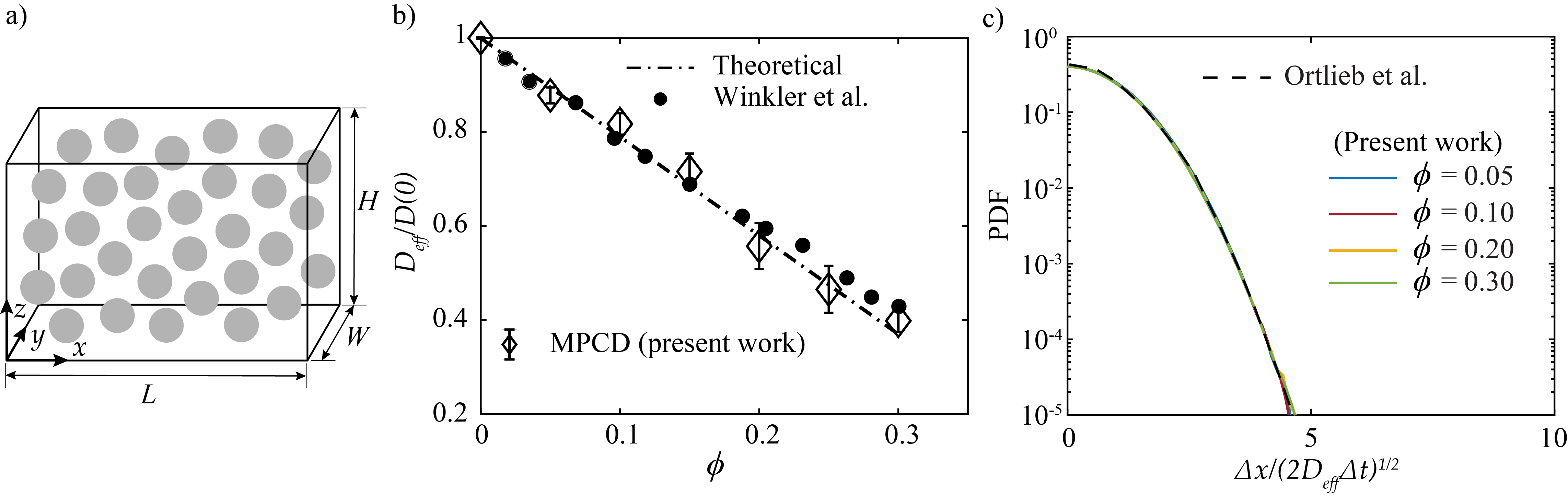}
        \put(55.5,25){\small \cite{Winkler2005}}
        \put(91,26.3){\small \cite{Ortlieb2019}}
    \end{overpic}
    \caption{Validations for colloidal suspension. a) the schematic of the simulation domain with spherical colloids (gray). b) Normalized translational diffusion coefficient of colloids ($D_{eff}/D(0)$) with respect to colloidal packing fraction ($\phi$). The filled black circles are from Winkler et al. \cite{Winkler2005}, the theoretical diffusion coefficient $D_{eff}/D(0) = 1-2.1\phi+O(\phi^2)$. c) Probability distribution function (PDF) of colloidal suspension with respect to the normalized displacement of the squirmer ($\Delta x/(2D_{eff}\Delta t)^{1/2}$) for different colloidal packing fractions ($\phi$). Dashed black line is from Ortlieb et al. \cite{Ortlieb2019}}
    \label{fig:Colloids_valid} 
\end{figure}

To verify the implementation of the colloidal suspension, we simulated multiple hard spheres of radius $R_c = 4a_0$ in a three-dimensional periodic domain with MPCD fluid. A schematic representation of the simulation domain is shown in Fig. \ref{fig:Colloids_valid}a. For the validation of monodisperse colloidal suspension, we compared the variation of the effective diffusion coefficients ($D_{eff}/D(0)$) of colloids with the colloidal packing fraction ($\phi = 0.05, 0.1, 0.2, \text{and} 0.3$) with Winkler et al. \cite{Winkler2005} (maximum error of $8.7\%$) and compared with the theoretical diffusion coefficient ($D_{eff}/D(0) = 1-2.1\phi+O(\phi^2)$) as shown in Fig. \ref{fig:Colloids_valid}b. The Fig. \ref{fig:Colloids_valid}b depicts the decrease in diffusion coefficient with an increase in the colloidal packing fraction. The considered MPCD parameters are as follows: domain size of $80a_0 \times 80a_0 \times 80a_0$, $N_c$ = 5, $\delta t = 0.1t_0$, $\delta t_s = 0.01t_0$, $k_BT = 1$, and $\epsilon = 1000$. Here, $t_0 = a_0\sqrt{m/k_BT}$ is the MPCD times scale. In Fig. \ref{fig:Colloids_valid}b, each numerical point is averaged over 5 realizations with each of $5\times10^6$ timesteps.

To validate the colloidal suspension, we also compared the Gaussianity of the colloidal displacement distribution with Ortlieb et al. \cite{Ortlieb2019} as shown in Fig. \ref{fig:Colloids_valid}c. We observe that the distribution of the displacement of the colloids in a Newtonian fluid follows the Gaussian distribution with kurtosis 3, irrespective of the colloidal packing fraction ($\phi = 0.05, 0.1, \text{and } 0.2$). For this study, we consider the following MPCD parameters: domain size of $80a_0$ cube, $N_c = 10$, $\delta t = 0.02t_0$, $\delta t_s = 0.002t_0$, $k_BT = 1$, and $\epsilon = 1000$.

\subsection{\label{sec:valid_sphere}Validations: Hard sphere between two parallel walls}

\begin{figure}[ht]
    \centering
    \begin{overpic}[width=0.6\linewidth]{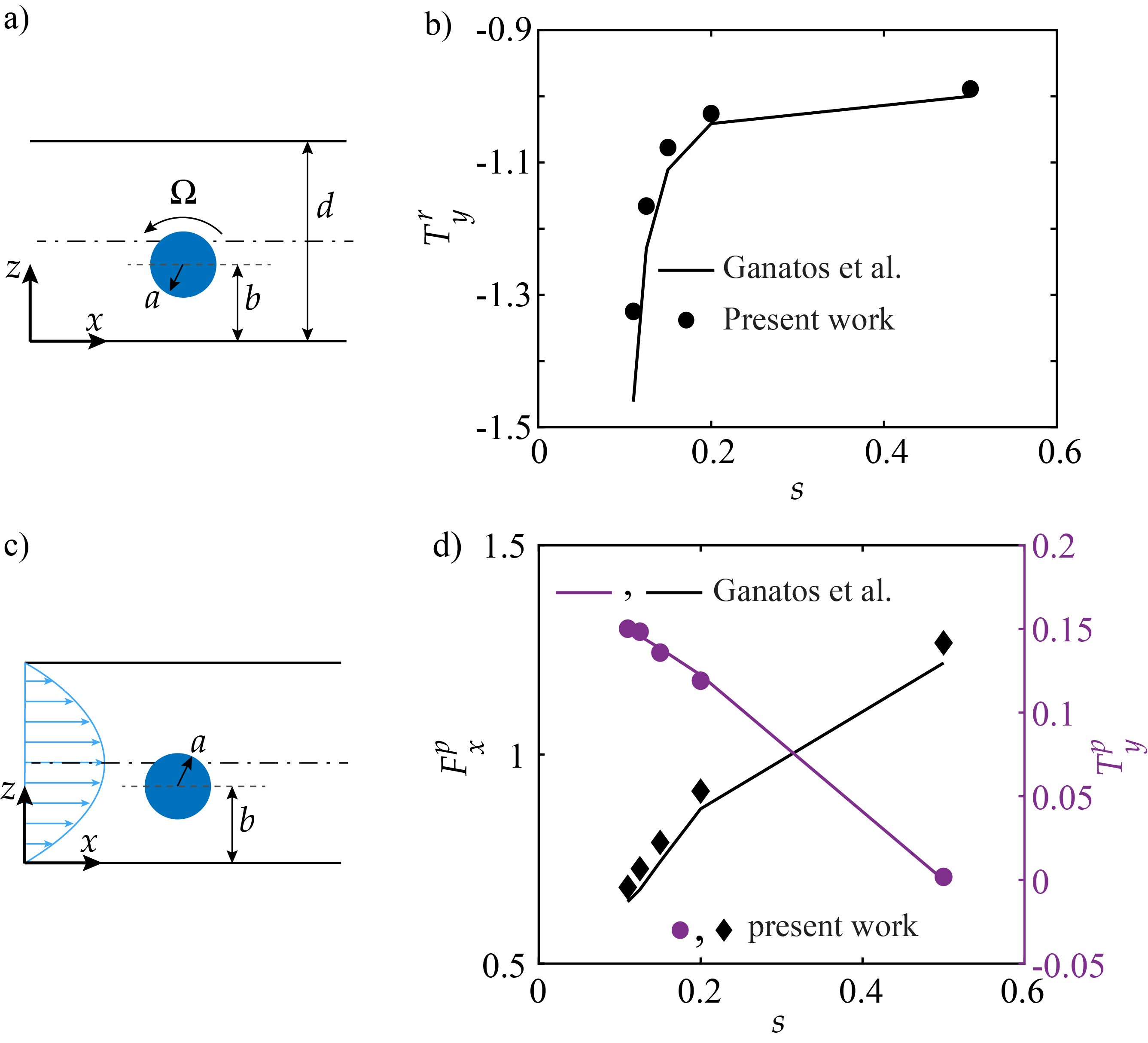}
        \put(79,66.5){\cite{Ganatos1980_part2}}
        \put(78,38.5){\cite{Ganatos1980_part2}}
    \end{overpic}
    \caption{a) Schematic representation of a sphere rotating with angular velocity ($\Omega$) about $y$ axis between two walls, b) Torque coefficient ($T^r_y$) of the hard sphere rotating along the $y$-axis between two parallel walls, c) Schematic representation of a sphere in a plane Poiseuille flow with maximum speed ($V_c$), and d) Force ($F^p_x$) and torque ($T^p_y$) coefficient of the hard sphere placed in a plane Poiseuille flow. (b, d) The thick lines are from Ganatos et al. \cite{Ganatos1980_part2}.}
    \label{fig:valid_G2}
\end{figure}

To check the hydrodynamic interactions between the wall and the sphere, we simulate a problem where a hard sphere is positioned between two walls and rotating with an angular velocity $\Omega$. We compared the torque coefficient ($T^r_y$) from Ganatos et al. \cite{Ganatos1980_part2}. A schematic representation of the simulation domain is shown in Fig. \ref{fig:valid_G2}a. We calculated the torque coefficient $T^r_y = T_{actual}/(8\pi\mu a^3 \Omega)$, where $T_{actual}$ is calculated from the angular momentum of the MPCD particles during streaming and collision steps, and $\mu$ represents the MPCD dynamic viscosity. We observed a significant comparison with Ganatos et al. \cite{Ganatos1980_part2}, as shown in Fig. \ref{fig:valid_G2}b, with an error range of 1-10\%. Here, $s = b/d$, $d$ is the distance between the two walls, and $b$ is the distance of the sphere center from the bottom wall. Considered MPCD parameters are: wall distance $40a_0$, $a = 4a_0$, $N_c = 30$, and $\delta t = 0.02t_0$.

To validate the implementation of the plane Poiseuille flow, we compare the force ($F^p_x$) and torque ($T^p_y$) coefficients from Ganatos et al. \cite{Ganatos1980_part2} for a sphere of radius a positioned in plane Poiseuille flow. Fig. \ref{fig:valid_G2}c depicts the schematic diagram of the inert hard sphere in a plane Poiseuille flow. We calculate the force and torque coefficients according to $F^p_x = F_{actual}/(6\pi\mu a u_m)$ and $T^p_y = T_{actual}/(8\pi\mu a^2 u_m)$, respectively. Here, $F_{actual}$ is calculated from the linear momentum of the MPCD particles during the streaming and collision steps, and $u_m$ is the maximum channel velocity. We observe a notable comparison, with an error ranging between $0.01\%$ and $7\%$, as shown in Fig. \ref{fig:valid_G2}d. The MPCD parameters are similar to the above problem.

\subsection{\label{sec:valid_qi}Validations: Squirmer in plane Poiseuille flow}

\begin{figure}[ht]
    \centering
    \includegraphics[width = \linewidth]{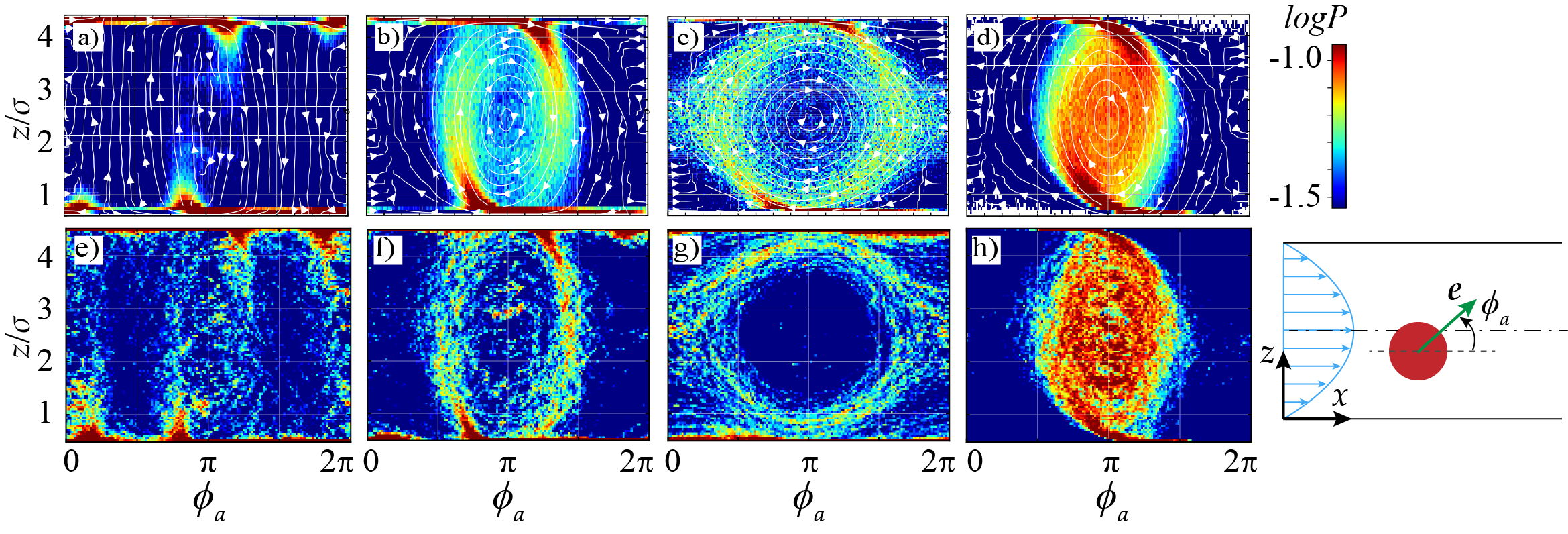}
    \caption{Validation of the squirmer in a plane Poiseuille flow. Right image depicts the schematic simulation domain with spherical squirmer orientation vector ($\bm{e}$), and azimuthal angle ($\phi_a$) in $xz$-plane. Joint probability distribution function of $z$ directional position and the azimuthal angle of the squirmer ($\phi_a$) for a) and e) $V_r = 0.28$, b) and f) $V_r = 2.25$, and c) and g) $V_r = 6.75$ with $\beta = -3$. and d) and h) $V_r = 2.25$ with $\beta = 3$. (a-d) are from Qi et al. \cite{Qi2020rheotaxis}, and (e-h) are from the present work. (a-d) Reproduced with permission from ref. \cite{Qi2020rheotaxis}. Copyright 2020 Physical Review Research under the terms of the  Creative Commons Attribution 4.0 International license [https://creativecommons.org/licenses/by/4.0/].}
    \label{fig:sq_flow}
\end{figure}

To validate the implementation of the interactions between the squirmer and the wall with flow conditions, we compared the joint probability distribution function ($P(z,\phi_a)$) of a single squirmer position in $z$ direction and azimuthal angle $\phi_a$ (the angle between the flow direction and squirmer orientation) with Qi et al. (Fig. \ref{fig:sq_flow}) \cite{Qi2020rheotaxis}. A schematic representation of the simulation domain is depicted on the right side of Fig. \ref{fig:sq_flow}. Here, $\sigma$ is the diameter of the squirmer. In Fig. \ref{fig:sq_flow}(a-d) are from the Qi et al. \cite{Qi2020rheotaxis} and (e-h) are from the present simulations. We compare the $P(z,\phi_a)$ for $\beta = -3$, $B_1 = 0.05$, and $V_r = 0.28$ (Fig. \ref{fig:sq_flow}(a,e)), 2.25 (Fig. \ref{fig:sq_flow}(b,f)), 6.75 (Fig. \ref{fig:sq_flow}(c,g)). We also compared the $P(z,\phi_a)$ for puller $\beta = 3$, $B_1 = 0.05$, and $V_r = 2.25$ (Fig. \ref{fig:sq_flow}(d,h)). From Fig. \ref{fig:sq_flow}, we observe a qualitatively similar high probability of the squirmer position near the wall with both upstream and downstream orientation, consistent with the finding of Qi et al. \cite{Qi2020rheotaxis}. In Fig. \ref{fig:sq_flow}, each contour is generated from the simulations of a minimum of 5 realizations with each $5\times10^6$ timesteps.

\section{\label{sec:without_flow}Confined squirmer in a quiescent colloidal suspension}

\begin{figure}[h]
    \centering
    \includegraphics[width = 0.65\linewidth]{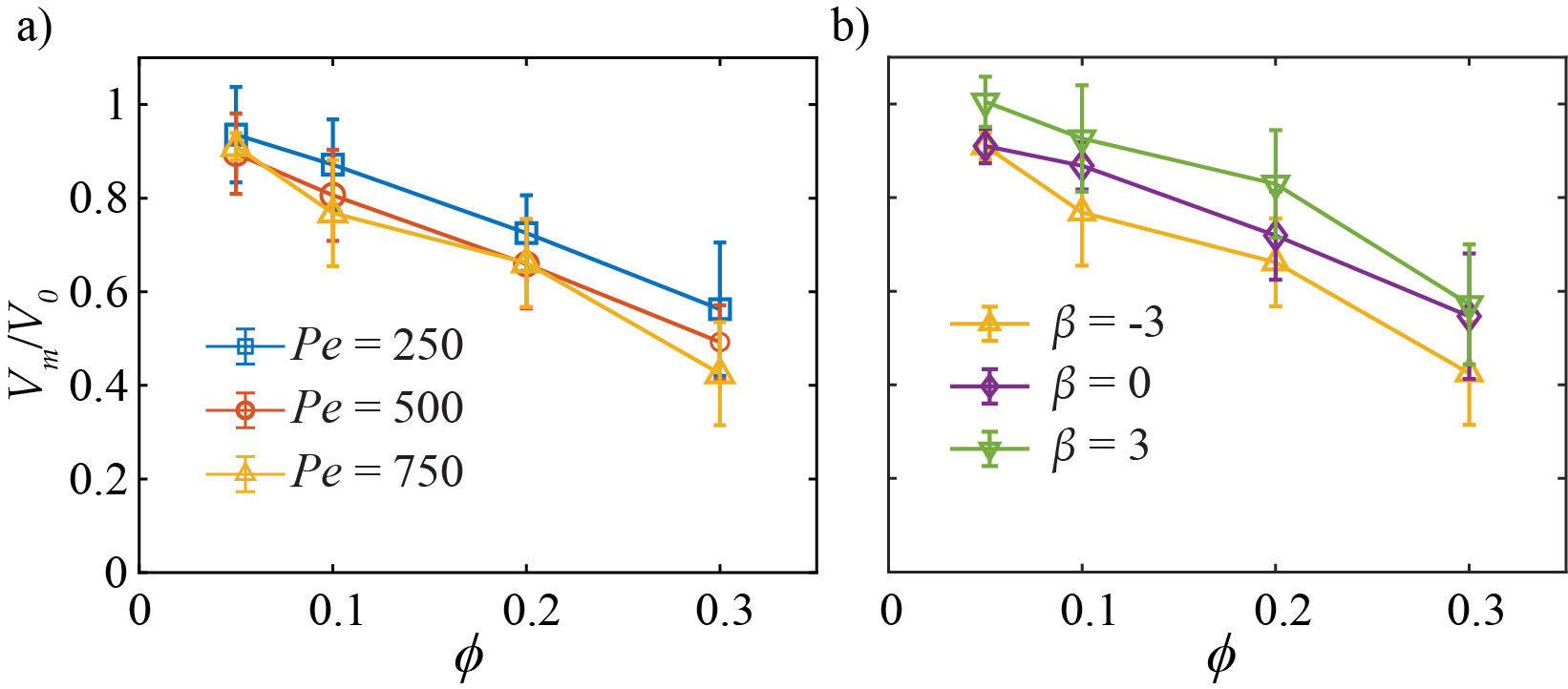}
    \caption{Normalized mean speed ($V_m/V_0$) of a) the pusher ($\beta = -3$) for different $Pe$ numbers, and b) different types of squirmers for $Pe = 750$ with respect to colloidal packing fraction ($\phi$) without flow. Each data point is an average over 10 realizations, each with $2\times10^5$ timesteps. The error bars show the standard deviation.}
    \label{fig:speed_sq_coll_wof}
\end{figure}

We simulated the motion of a single squirmer in a colloidal suspension between two parallel walls to analyze the mean speed ($V_m$) of different types of squirmers in various colloidal packing fractions ($\phi$). From Fig. \ref{fig:speed_sq_coll_wof}(a,b), which shows the normalized mean speed of the squirmer ($V_m$) by the theoretical squirmer speed ($V_0$) with respect to $\phi$, we observe the decrease in the squirmer mean speed with an increase in $\phi$, as we expected. However, the rate of decrease in the squirmer mean speed is higher for high $Pe$ number compared to the low $Pe$ number in the case of a pusher ($\beta = -3$), as shown in Fig. \ref{fig:speed_sq_coll_wof}a. Therefore, the higher the thermal fluctuations, the higher the squirmer mean speed for a constant $\phi$, where the high thermal fluctuations reduce the local trapping of the squirmer due to the hydrodynamic attraction forces near a colloid \cite{Ramprasad2025} or wall. The role of thermal fluctuations in allowing the bacteria to escape a hydrodynamic trap is illustrated by Drescher et al. \cite{drescherPNAS2011}. From Fig. \ref{fig:speed_sq_coll_wof}b, we observe that for a constant $\phi$ and $Pe = 750$, the puller moves with a higher speed compared to the neutral squirmer, which moves faster than the pusher. The high speed of the neutral- and puller-type squirmer compared to the pusher is due to the hydrodynamic repulsion forces between the colloid and the neutral or puller-type squirmer \cite{Lauga2009TheMicroorganisms, Gotze2010}, which causes the squirmer to avoid trapping near the colloid \cite{Ramprasad2025} and wall.

\section{Effect of flow strength with colloids}
\label{asec:flow_strength_effect_wc}

This section presents a detailed study of the effects of the squirmer parameter and flow strength in the colloidal suspension case for $Pe = 750$. The simulation model and parameters are identical to those specified in the main script, unless otherwise specified separately.

\subsection{Translational and directional behaviors of squirmer at high flow strength}
\label{assec:traj_Vr4}

\begin{figure}[ht]
    \centering
    \includegraphics[width = 0.8\linewidth]{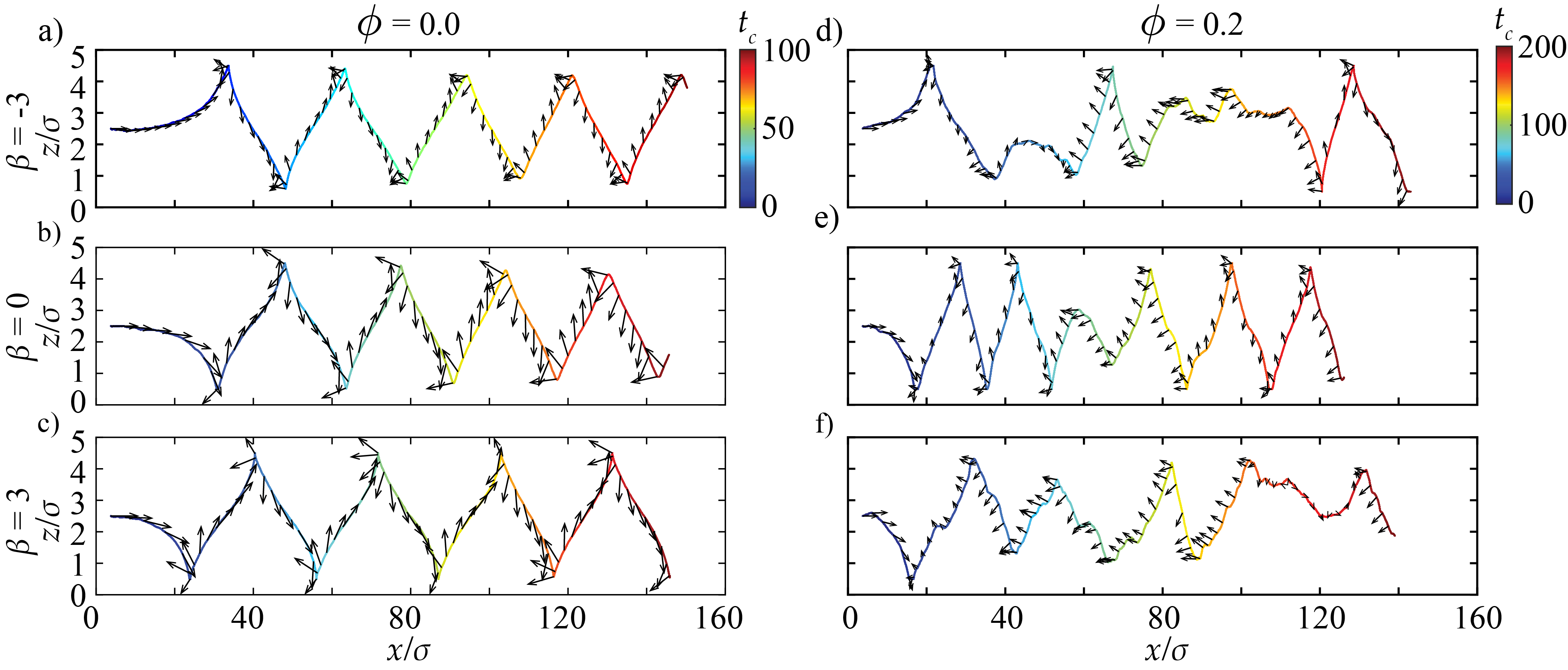}
    \caption{Trajectory of the (a, d) pusher ($\beta = -3$), (b, e) neutral ($\beta = 0$), and (c, f) puller ($\beta = 3$) squirmer for (a-c) $\phi = 0$, (d-f) $\phi = 0.2$, and $V_r = 4$. The colorbar indicates the characteristic time of the squirmer.}
    \label{fig:traj_Pe750_ball_phi00_20_ub_40_ds_n}
\end{figure}

Here, we illustrate the squirmer's translational and directional behaviors in plane Poiseuille flow, without ($\phi = 0$) and with ($\phi = 0.2$) colloids, for a flow strength of $V_r = 4$. The simulation method and parameters are identical to those specified in the main script, unless otherwise specified. Here, the Fig. \ref{fig:traj_Pe750_ball_phi00_20_ub_40_ds_n} shows the trajectories of the squirmers with instantaneous squirmer orientations. We observe that at high flow strength ($V_r = 4$), all types of squirmers (pushers, neutral, and pullers) oscillate between two walls with a dominating cross-streaming (Fig. \ref{fig:traj_Pe750_ball_phi00_20_ub_40_ds_n}(a-c)). The oscillatory motion of the squirmer arises from the interplay between shear-induced rotation in the Poiseuille flow, self-propulsion along the body axis, and wall-induced hydrodynamic interactions, leading to a stable cross-stream limit cycle \cite{Zottl2012}. However, in the presence of colloids, we observe an increase in the upstream orientation of the squirmers (Fig. \ref{fig:traj_Pe750_ball_phi00_20_ub_40_ds_n}(d-f)) along with the reduction in the amplitude of the oscillations. This shows that the squirmers are more confined to the channel center at high flow strength in the presence of the colloids.

\subsection{Spacial distribution of the time-averaged streamwise squirmer velocity}
\label{assec:velocity}

\begin{figure}[h]
    \centering
    \includegraphics[width = 0.9\linewidth]{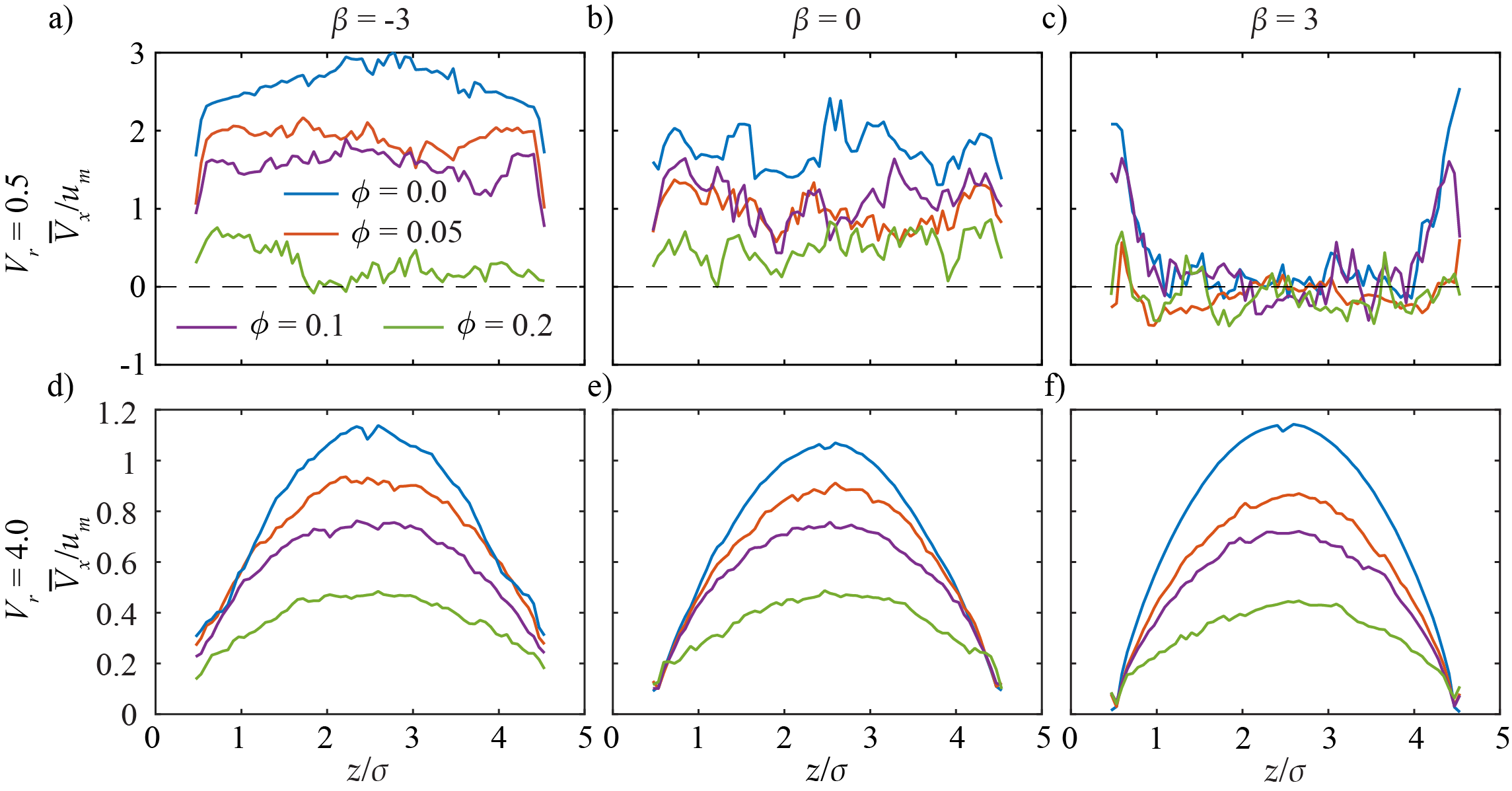}
    \caption{Local averaged velocity of (a, d) pusher ($\beta = -3$), (b, e) neutral ($\beta = 0$), and (c, f) puller ($\beta = 3$) in the flow direction with respect to the velocity gradient direction at flow strength of (a-c) $V_r = 0.5$ and (d-f) $V_r = 4$ for different colloidal packing fractions ($\phi$). Here, the local flow directional velocity of the squirmer ($\overline{V}_x$) is normalized by the maximum flow speed ($u_m$) of the plane Poiseuille flow.}
    \label{fig:v_distrbution_Pe750_ball_ubU05_4_vx}
\end{figure}

With the increase in colloidal packing fraction ($\phi$), the mobility of the squirmer decreases due to the presence of colloids in the system, along with the tendency of the squirmer to adopt upstream-oriented swimming. From Fig. \ref{fig:v_distrbution_Pe750_ball_ubU05_4_vx}, we observe that the local streamwise velocity of the squirmer ($\overline{V}_x$) decreases with increasing $\phi$. In particular, at low flow strength ($V_r = 0.5$), the pusher and neutral squirmers exhibit near-zero local velocity along the flow direction in the central region of the channel at high packing fraction ($\phi = 0.2$), indicating a strong suppression of downstream transport (Fig. \ref{fig:v_distrbution_Pe750_ball_ubU05_4_vx}(a, b)). In contrast, for pullers (Fig. \ref{fig:v_distrbution_Pe750_ball_ubU05_4_vx}c), we observe a significant negative local velocity along the flow direction, indicating dominant upstream swimming. This behavior is consistent with their orientational preference, which leads to a reduced or reversed projection of the swimming velocity onto the flow direction. Additionally, the observed fluctuations in the local averaged velocity profiles arise from the presence of colloids and the dominance of stochastic effects at low flow strength ($V_r = 0.5$), which introduce variability in the squirmer trajectories and orientations.

At high flow strength ($V_r = 4$), the local streamwise velocity profiles of the squirmer follow the background parabolic shape of plane Poiseuille flow (Fig. \ref{fig:v_distrbution_Pe750_ball_ubU05_4_vx}(d-f)). We observe a reduction in ($\overline{V}_x$) with increasing $\phi$ along the velocity gradient direction. Differences due to the squirmer type ($\beta$) persist but are less pronounced. We observed that the pushers maintain comparatively higher velocities than the neutral and pullers, consistent with their tendency to align more favorably with the flow direction. In contrast, neutral and pullers exhibit slightly reduced velocities, particularly near the walls, where their orientations deviate from the flow direction. The fluctuations in the local velocity profiles are also significantly diminished, reflecting the reduced influence of stochastic effects under strong advective transport.

\bibliography{ref.bib}

\end{document}